\documentclass[12pt]{iopart}
\usepackage{amssymb}
\usepackage[dvips]{color}
\usepackage{graphicx}
\usepackage{bm}

\begin{document}
\title{Gravastars must have anisotropic pressures}

\author{%
Celine Cattoen \footnote[1]{celine.cattoen@mcs.vuw.ac.nz},
Tristan Faber \footnote[2]{tristan.faber@mcs.vuw.ac.nz},
and
Matt Visser \footnote[3]{matt.visser@mcs.vuw.ac.nz}
}
\address{School of Mathematics, Statistics, and Computer Science, \\
Victoria University of Wellington, \\
P.O.Box 600, Wellington, New Zealand}


\begin{abstract}


One of the very small number of serious alternatives to the usual concept of an astrophysical
black hole is the ``gravastar'' model developed by Mazur and Mottola; and a related phase-transition model due to Laughlin \emph{et al}. We consider a generalized class of similar models that exhibit continuous pressure --- without the presence of infinitesimally thin shells. By considering the usual TOV equation for static solutions with negative central pressure, we find that gravastars cannot be perfect fluids --- anisotropic pressures in the ``crust'' of a gravastar-like object are unavoidable. The anisotropic TOV equation can then be used to bound the pressure anisotropy.
The transverse stresses that support a gravastar permit a higher compactness than is given by the Buchdahl--Bondi bound for perfect fluid stars. Finally we comment on the qualitative features of the equation of state that gravastar material must have if it is to do the desired job of preventing horizon formation.

\bigskip

{gr-qc/0505137;  

Latest version: 27 May 2005;  \LaTeX-ed \today}

\end{abstract}
\pacs{04.20.-q; 04.20.Cv}

\maketitle
\def\d{{\mathrm{d}}}
\def\rr{\tilde{r}}
\def\rhoavg{\bar{\rho}}

\section{Introduction}

Although the concept of a black hole is well-established and generally accepted in the relativity, astrophysics, and particle physics communities, one sometimes encounters  a certain amount of scepticism regarding the physical reality of the mathematical solution, and wariness regarding the interpretation of observational data~\cite{Marek}. The simplicity of the Schwarzschild solution comes at the expense of a central singularity, and an event horizon at the Schwarzschild radius $R_\mathrm{Schwarzschild}=2M$. Critics of black holes, at least the rational ones, seek an alternative configuration of matter that concentrates as much energy density as possible within a radius of $R \gtrsim 2M$ while avoiding the formation of the singularity and the event horizon.

One of the small number of serious challenges to the usual concept of black holes is the ``gravastar'' (\textit{gra}vitational \textit{va}cuum \textit{star}) model that was recently developed by Mazur and Mottola ~\cite{MM1, MM2, MM3}. In the gravastar picture, or the very closely related quantum phase transition picture developed by Laughlin \emph{et al.}~\cite{Laughlin1, Laughlin2}, the quantum vacuum undergoes a phase transition at or near $R_\mathrm{Schwarzschild}$ where the event horizon would have been expected to form. 
Somewhat related models, differing in motivation and technical details, can be traced back to Gliner~\cite{Gliner} and Dymnikova~\cite{Dymnikova}.

In the Mazur--Mottola model, a suitable segment of de Sitter space (with an equation of state $\rho = -p>0$) is chosen for the interior of the compact object while the outer region of the gravastar consists of a (relatively thin) finite-thickness shell of stiff matter ($p=\rho$) that is in turn surrounded by Schwarzschild vacuum ($p=\rho=0$).\footnote{Because of the negative pressure core the gravastar exhibits some of the properties of the ``tension stars'' considered by Katz and Lynden--Bell~\cite{Lynden-Bell}, and by Comer and Katz~\cite{Comer}, though the intent and scope of the model is rather different.}
Apart from these three explicitly mentioned layers, the Mazur--Mottola model requires two additional infinitesimally-thin shells with surface densities $\sigma_\pm$, and surface tensions $\vartheta_\pm$, that compensate the discontinuities in the pressure profile and stabilize this 5-layer onion-like construction, effectively by introducing delta-function anisotropic pressures \cite{MM1,MM2,MM3}. Since infinitesimally thin shells are a mathematical abstraction,  for physical reasons it is useful to minimize the use of thin shells, either by successfully reducing the system to a 3-layer onion with \emph{one} thin shell surrounded by segments of Schwarzschild and de Sitter space~\cite{visser04}, or more boldly (as in this article) by attempting to replace the thin shells completely with a continuous layer of finite thickness.\footnote{See also reference~\cite{Volovik} for related ideas.}

Indeed, in the present article we demonstrate that the pressure anisotropy implicit in the Mazur--Mottola infinitesimally thin shell  is not an accident, but instead a \emph{necessity} for all gravastar-like objects. That is, attempting to build a gravastar completely out of perfect fluid will always fail. (Either the gravastar swells up to infinite size, or a horizon will form despite one's best efforts, or worse a naked singularity will manifest itself.)  We derive this result by working with configurations where pressure is assumed continuous and differentiable, and analyzing the resulting static geometry using first the \emph{isotropic}
 TOV equation, and then the \emph{anisotropic} TOV equation.

\section{The geometry}

We adopt coordinates that allow us to write any static spherically symmetric geometry in the form
\begin{equation}\label{eq:metric}
\d s^{2}= 
-\exp \left[ -2\int_r^\infty g(\rr)\, \d\rr \right] \d t^{2}
+\frac{\d r^{2}}{1-2m(r)/r}
+r^{2}\; \d\Omega^{2} \, .
\end{equation} 
We chose the two metric functions so that $g(r)$ represents the locally measured gravitational acceleration, which is pointing inwards for positive $g(r)$, and so that $m(r)$ is the total mass-energy confined in a sphere with radius $r$. This interpretation can be justified by invoking the Einstein equations for the static stress-energy tensor $T^\alpha_{~\beta} = \mathrm{diag} [ -\rho, p_r, p_t, p_t ]$. The $tt$-field equation then yields
\begin{equation}
\label{eq:mass}
m(r)=4\pi\int_0^r \rho(\rr)\,\rr^2\,\d\rr
\end{equation} 
and the $rr$-field equation gives
\begin{equation}
\label{eq:grav}
g(r)=\frac{m(r)+4\pi p_r(r)\, r^3}{r^2 \left[ 1-2m(r)/r \right] }
=\frac{4\pi\,r}{3}\, \frac{\rhoavg(r)+3p_r(r)}{1-2m(r)/r}
\end{equation} 
where we made use of the average density $\rhoavg(r)\equiv m(r)/(\frac{4}{3}\pi r^3)$.

From now on, in the interests of legibility, we discontinue indicating the explicit $r$-dependence of all relevant functions. The remaining field equation for the transverse pressure is quite messy. Instead, we make use of the Bianchi identities and replace it with the covariant conservation of stress energy:
\begin{equation}
\label{eq:pre-TOV}
\frac{\d p_r}{\d r} = -(\rho+p_r)\, g + \frac{2\left( p_t - p_r \right)}{r}.
\end{equation}
In the case of isotropic pressures $p=p_r=p_t$ this leads to the standard TOV equation, which can be written in any of the equivalent forms
\begin{equation}
\label{eq:TOV}
\frac{\d p}{\d r} = -(\rho+p)\, g = 
-\frac{(\rho + p) \; (m+4\pi p\, r^3)}{r^2 \left[ 1-2m(r)/r \right] } 
= -\frac{4\pi\,r}{3}\, \frac{(\rho+p)\;(\rhoavg+3p)}{1-2m(r)/r}.
\end{equation}
For anisotropic pressures we find it convenient to define the dimensionless anisotropy parameter
\begin{equation}
\Delta \equiv {p_t - p_r \over \rho},
\end{equation}
so that in terms of this parameter the anisotropic TOV becomes
\begin{equation}
\label{eq:TOV2}
\frac{\d p_r}{\d r}  = -\frac{4\pi\,r}{3}\, \frac{(\rho+p_r)\;(\rhoavg+3p_r)}{1-2m(r)/r} 
+{2\;\rho\;\Delta\over r}.
\end{equation}
Before going on to explore some of the key features and consequences of equations \eref{eq:TOV} and  \eref{eq:TOV2}, we must define the class of spacetime geometries that we are particularly interested in.

\section{Key features of smooth gravastar models}

\begin{figure}[htb]
\begin{center}
\input{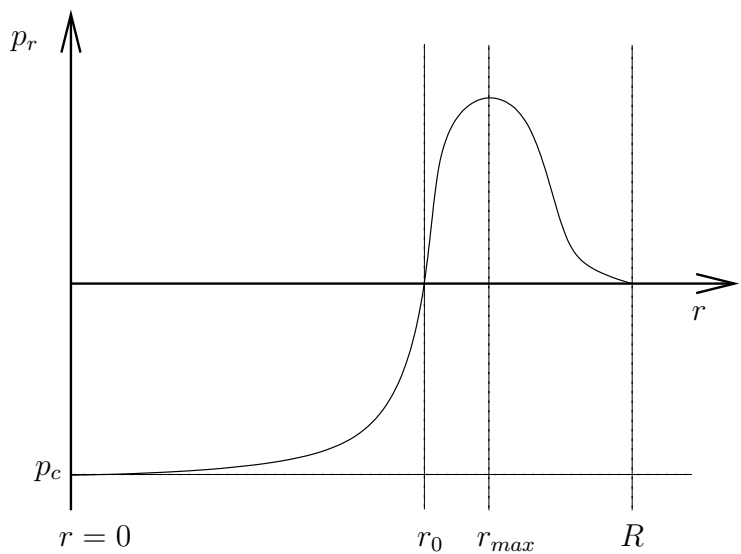}
\end{center}
\caption{\label{F:pressure}
Qualitative sketch of radial pressure as a function of $r$ for a gravastar .}
\end{figure}

To have a useful model, we should retain as much of standard physics as possible,  while introducing a minimum of ``new physics''. In the spirit of Mazur and Mottola~\cite{MM1,MM2,MM3}, and Laughlin \emph{et al.}~\cite{Laughlin1, Laughlin2}, we will keep the density positive throughout the configuration but permit the pressure to become negative in the gravastar interior.  To avoid infinitesimally thin shells one must then demand that the radial pressure $p_r$ is continuous (though the density need not be continuous, and typically is not continuous at the surface of the gravastar). Qualitatively the radial pressure is taken to be that of figure \ref{F:pressure}.

That is, to ``smooth out''  the infinitesimally thin shells of the Mazur--Mottola gravastar model, we shall consider static spherically symmetric geometries such that:
\begin{itemize}
\item Inside the gravastar, $ r < R$, the density is everywhere positive and finite.

\item The  central pressure is negative, $p_c<0$, and in fact $p_c = - \rho_c$.
\\
(We do \emph{not} demand $\rho=-p_r=-p_t$ except at the centre.)

\item The spacetime is assumed to \emph{not} possess an event horizon. \\
This implies that $\forall r$ we have $2m(r)<r$.

\end{itemize}
These three features, positive density, negative central pressure, and the absence of horizons, are the three most important features characterizing a gravastar. Other important features are:
\begin{itemize}
\item To keep the centre of the spacetime regular, we enforce both $p_r'(0)=0$ 
and $p_c = p_r(0)=p_t(0)$.

\item There should be a pressure maximum in the general vicinity of the Schwarzschild radius, $r_\mathrm{max}\approx~R_\mathrm{Schwarzschild}$, satisfying  $p_r(r_\mathrm{max})>0$, and $p_r'(r_\mathrm{max})=0$. \\
(This permits the physics in the region $r\gg r_\mathrm{max}$ to be more or less standard.)

\item There should be exactly two radii where the radial pressure vanishes:
\begin{itemize}
\item  The first pressure zero $p_r(r_0)=0$, where $p_r'(r_0) > 0$, and 
\item
the second pressure zero $p_r(R)=0$, where $p_r'(R)\leq 0$. \\
The point $R$,  (which by construction must satisfy $R>r_\mathrm{max}>r_0$), is called the surface of the gravastar.
\end{itemize}

\item The pressure profile $p_r(r)$ should be continuous.\\
 (In contrast, it is sometimes useful to  allow $p_t(r)$ to be discontinuous.)
 
\item The strong energy condition [SEC;  $\rho+p_r+2p_t\geq 0$] is definitely violated, at least near the centre of the gravastar.

\item We choose to enforce the null energy condition [NEC; $\rho+p_i\geq 0$] throughout the gravastar. In view of our first comment that density is everywhere positive, this implies that we are enforcing the  weak energy condition [WEC; $\rho+p_i\geq 0$ and $\rho\geq 0$]. 

\item We impose no restriction regarding the dominant energy condition [DEC; $\rho\geq0$ and $|p_i|\leq \rho$], and in fact we shall see that the DEC must fail in parts of the gravastar that are sufficiently ``close'' to forming a horizon.
\end{itemize}

We comment that the Mazur--Mottola model might then be viewed as the  limiting case where $r_\mathrm{max}\to r_0$, while $\rho=-p_r=-p_t$ is strictly enforced for $r<r_0$, and where an additional thin shell is placed at the surface [so that $p_r(R^-)>0$].  The 3-layer variant considered in~\cite{visser04} is effectively the Mazur--Mottola model subject to the additional limit $R\to r_\mathrm{max}$, so in this model $r_0=r_\mathrm{max}=R$ and there is a single thin shell at $r_\mathrm{max}$ with de Sitter geometry inside and Schwarzschild geometry outside.

Somewhat similarly, the Laughlin \emph{et al} model may be viewed as the singular double limit $r_0\to R_\mathrm{Schwarzschild}$ from below, while $R\to R_\mathrm{Schwarzschild}$ from above, and with $\rho=-p_r=-p_t$ strictly enforced for $r<R_\mathrm{Schwarzschild}$. There is implicitly a singular infinitesimally thin shell located exactly at $R_\mathrm{Schwarzschild}$ with infinite surface tension.

The Gliner~\cite{Gliner} and Dymnikova~\cite{Dymnikova} proposals all satisfy the constraint  $\rho=-p_r$ everywhere throughout the configuration. For the purposes of this article we view this as an unnecessary specialization.

One might wonder why we do not place the gravastar surface at $r_0$? After all the pressure is by definition zero there so we can smoothly join it on to an exterior Schwarzschild solution.  The reason for not doing so is a purely pragmatic one based on the fate of infalling positive-pressure matter. If we start without a positive-pressure region of type $(r_0,R)$, then any infalling positive-pressure  matter that accumulates above $r_0$ will automatically generate a positive-pressure region of type $(r_0,R)$. The only way to avoid a positive-pressure region of type $(r_0,R)$ is if the negative-pressure matter in the $(0,r_0)$ region immediately catalyzes any infalling positive-pressure matter into negative-pressure matter.  This scenario has its own risks, and for now we will keep the negative-pressure matter deep in the core, discretely hidden behind a layer of positive-pressure matter. (This is exactly what Mazur and Mottola did with their layer of stiff matter.)

\section{Failure of isotropic pressure in the gravastar ``crust''}
\label{sec:iso_fail}

For the time being, we are only considering perfect fluids, so that the pressure is isotropic: $p= p_r= p_t$  and $\Delta=0$ throughout this section.
Consider first the isotropic TOV equation \eref{eq:TOV} at $r=r_0$ where the pressure is first zero.  We find that
\begin{equation}
\left. \frac{\d p}{\d r}\right|_{r=r_0}  = - \frac{4\pi r_0}{3} \frac{\rho\,\rhoavg}{1-2m/r_0}.
\end{equation}
But the LHS is by assumption positive, while the RHS is by assumption negative. Therefore the isotropic TOV \emph{cannot} hold at the point $r_0$.

Secondly, consider the point of maximal positive pressure, $r=r_\mathrm{max}$. It follows from the isotropic TOV equation  (\ref{eq:TOV}) that
\begin{equation}
\left. \frac{\d p}{\d r}\right|_{r=r_\mathrm{max}}  =
 - \frac{4\pi r_\mathrm{max}}{3} \;
   \frac{(\rho+p)(\rhoavg+3p)}{1-2m/r_\mathrm{max}}.
\end{equation}
But $\rho>0$ everywhere inside the gravastar, and by assumption $p(r_\mathrm{max})>0$ in the class of models we consider. So the LHS is zero while the RHS is negative. Therefore the isotropic TOV \emph{cannot} hold at the point $r_\mathrm{max}$.

In fact the objects under investigation have an increasing (radial) pressure for the entire range $r_0 \leq r < r_\mathrm{max}$ while the (radial) pressure in the same interval is positive. That is the LHS of the isotropic TOV is positive in this region, while the RHS of the isotropic TOV is negative. Hence, it follows by contradiction from (\ref{eq:TOV}) that isotropic pressure in that interval is not able to satisfy the TOV equation, and thus a static spacetime geometry can only be obtained with anisotropic pressures.

The same argument holds for a larger interval that extends below $r_0$ and into the negative pressure region: Assuming the NEC, we have $\rho+p\geq 0$. Therefore, by equation \eref{eq:TOV}, we must have $\rhoavg+3p<0$ if the pressure gradient for isotropic pressures is to be positive.  Now at the centre of the gravastar $(\rhoavg+3p)_c = \rho_c + 3 p_c = - 2\rho_c < 0$, while at $r_0$ we have $(\rhoavg+3p)_0 = \rhoavg_0 > 0$. Therefore   $\rhoavg+3p$ changes sign somewhere in the interval $r\in(0,r_0)$. 
Define $r_g$ to be the location where $\rhoavg+3p$ changes sign. We conclude that pressure isotropy fails for the entire region where both $\rhoavg+3p>0$ and $p'\geq0$, and quite possibly fails for an even larger region. That is, isotropy definitely fails on the region $r_g < r \leq r_\mathrm{max}$, a region which definitely encompasses $r_0\leq r \leq r_\mathrm{max}$. 

\begin{figure}[htb]
\begin{center}
\input{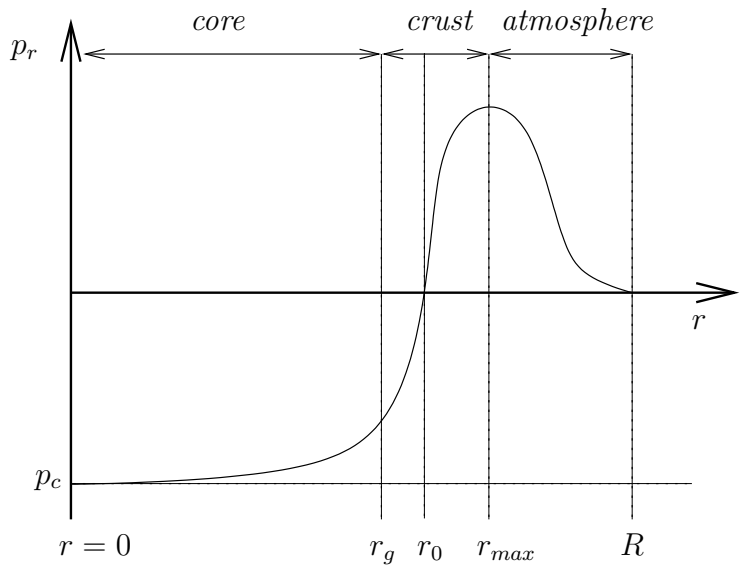}
\end{center}
\caption{\label{F:crust}
Qualitative sketch of  gravastar labelling the ``core'', ``crust'', and ``atmosphere''.}
\end{figure}

Let us call the region $r\in(0,r_g)$ the ``core'' where the physics is qualitatively (if not necessarily quantitatively) similar to that of de Sitter space. In particular in the ``core'' the local acceleration due to gravity [given by  equation \eref{eq:grav}] is outward.  Similarly let us call $r\in(r_g,r_\mathrm{max})$  the ``crust'', where physics is still definitely ``unusual''.  In the ``crust'' the local acceleration due to gravity is inward, but the pressure still rises as one moves outward. Finally, call $r\in(r_\mathrm{max}, R)$ the ``atmosphere'', where the physics is ``normal'', or rather as normal as it is going to get in a gravastar. In the ``atmosphere'' the local acceleration due to gravity is inwards, and the pressure decreases as one moves outwards. (See figure \ref{F:crust}.) With these definitions, we see that pressure is guaranteed to be anisotropic throughout the ``crust''. Note that even if we were to dispense with the entire positive-pressure region by chopping the gravastar off at $r_0$, there is still an anisotropic crust in the region $(r_g,r_0]$.

To conclude this section, we summarize that a static spherically symmetric object with positive density, negative central pressure and vanishing pressure at the surface cannot be supported by isotropic pressures alone --- there are no perfect fluid gravastars.

\section{Schwarzschild interior solution and the Buchdahl--Bondi bound}

\begin{figure}[htb]
\begin{center}
\input{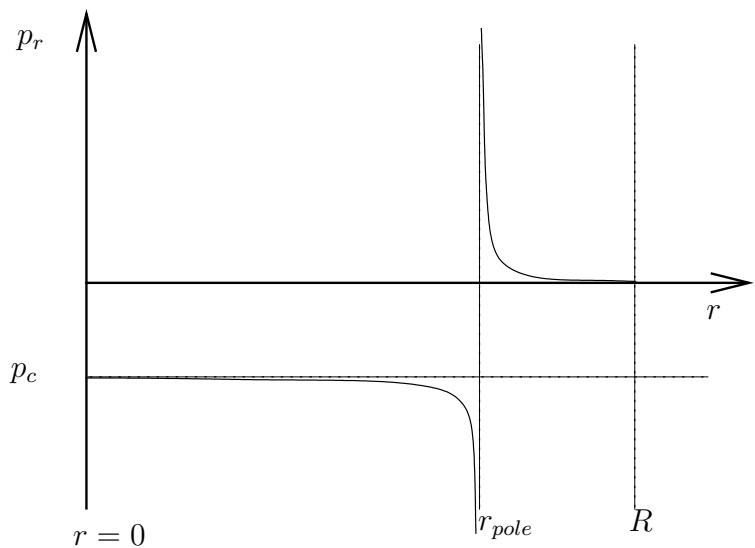}
\end{center}
\caption{\label{F:pole}
Qualitative sketch of pressure pole in the interior 
Schwarzschild solution for $2M/R > 8/9$.}
\end{figure}

A particular illustration of what goes wrong when one tries to build a perfect fluid gravastar is provided by the Schwarzschild interior solution. In this model one assumes constant positive density $\rho_*$ throughout the whole star and invokes the isotropic TOV equation \eref{eq:TOV}. Despite common misconception, this does \emph{not} mean the star is incompressible or that the speed of sound goes to infinity. See for instance \cite[p. 609 ff]{MTW} and \cite{Rahman}. In this situation one has the well-known analytical solution~\cite{MTW}
\begin{equation}\label{eq:int_SS}
p_*(r) = \rho_*\;
 \frac{\sqrt{1-2m_*(r)/r}-\sqrt{1-2M/R}}{3\sqrt{1-2M/R}-\sqrt{1-2m_*(r)/r}}
\end{equation} 
where $M$ and $R$ are the total mass and surface radius and $m_*(r) \equiv M (r/R)^3$. The central pressure
\begin{equation}
p_c^* = \rho_* \frac{1-\sqrt{1-2M/R}}{3\sqrt{1-2M/R}-1}
\end{equation}
certainly diverges for $2M/R \rightarrow 8/9$ and is positive for all $2M/R < 8/9$. This is a first indication of the existence of the Buchdahl--Bondi bound, but by itself is not enough to derive this bound for arbitrary perfect fluid spheres \cite{Martin}. Since we are only interested in this model as an example, we need not worry about fully general statements and simply note that
\begin{equation}\label{eq:neg_pcstar}
p_c^* < 0 \qquad \Longleftrightarrow  \qquad \frac{8}{9} \: < \: \frac{2M}{R} \: < \: 1 \, .
\end{equation}
While this is a perfectly sensible solution in the mathematical sense one would generally rule it out physically because of the negative central pressure. If we leave our prejudice against negative pressures aside, adopting the gravastar philosophy, we find that with the choice of \eref{eq:neg_pcstar}, the pressure profile \eref{eq:int_SS} will have a first order pole at
\begin{equation}
r_\mathrm{pole} = 3R\sqrt{1-\frac{8/9}{2M/R}} \: < \: R \, .
\end{equation} 
Note that as $2M/R$ goes from 8/9 to 1, the position of this pole moves from the centre of the star to the surface of the star.
The situation is qualitatively sketched in figure \ref{F:pole}. Note that the NEC must be violated sufficiently close to the pole.

This is now clearly unphysical, even if one is willing to accept negative pressures and even violations of the NEC. It is unphysical for more prosaic reasons because the pressure pole implies a curvature singularity --- in fact the $R_{\hat t\hat r\hat t\hat r}$ and $R_{\hat t\hat\theta\hat t\hat\theta}$ orthonormal components of the Riemann tensor are infinite, so that one has a naked singularity. 
The reason why we mention this specific example is because we shall soon see this ``pressure pole'' behaviour is generic --- continuous solutions with isotropic pressure are not possible as we have already shown in section \ref{sec:iso_fail}.\footnote{Consider for example reference~\cite{Bilic}, where a perfect fluid Chaplygin gas $\rho\propto 1/p$ is considered. The surface of their configuration occurs at $\rho=0$ where $p=-\infty$, at least as one approaches the surface from below. So the surface of their configuration is a naked singularity, in agreement with the observations above.}

\section{The fate of a negative pressure perfect fluid sphere}

Consider now an arbitrary gravastar (of the type defined above). 
We note that the ``compactness'' $\chi\equiv2m(r)/r$ satisfies
\begin{equation}
\fl
\left[{2m(r)\over r}\right]' = 8\pi\; \rho \;r - {2m\over r^2} 
= {8\pi\;r\over3} \; [3\rho-\rhoavg] = {8\pi\;r\over3}\; [3(\rho+p) -(\rhoavg+3p)].
\end{equation}
But the first term on the RHS is non-negative by the NEC, while the second term is by definition negative on $[0,r_g)$, so the compactness $2m(r)/r$ is monotone increasing on the range $[0,r_g)$. This is more unusual than one might expect from the simplicity of the argument. While the compactness of a normal perfect fluid star tends to increase as one moves outwards, in a normal star it  also can be subject to oscillations that make the overall picture quite subtle~\cite{Harrison, Yunes}.

Consider now a perfect fluid sphere with negative central pressure that satisfies the NEC. Since we have already seen that isotropy is violated on $(r_g,r_\mathrm{max})$, the only way we can maintain the perfect fluid nature of the sphere is if $r_g\to\infty$. (Whence also $r_0\to\infty$.) But since we do not want a horizon to form, the compactness must be bounded above by unity. And since we have just shown the compactness is monotonic we now see
\begin{equation}
\lim_{r\to\infty} {2m(r)\over r} = \chi_*; \qquad \chi_*\in(0,1].
\end{equation}
So not only does a NEC-satisfying perfect-fluid  gravastar expand to infinite volume, it also has infinite mass.

To avoid the physical size of the gravastar blowing up to $r_0\to\infty$, our options now are rather limited: We could permit the development of a horizon at finite $r$, which defeats the whole point of the exercise,  or we could permit something even worse.  If we permit NEC violations then it is possible to arrange for the  development of a pressure pole at finite $r<r_0$. To see how this is possible, consider  the configuration
\begin{equation}
p(r) \approx {\Gamma\over r-r_p},
\end{equation}
and let us check that it
is compatible with the \emph{isotropic} TOV equation. First for $r<r_p$ the pressure is negative, and for $r>r_p$ the pressure is positive, which is appropriate for a modified gravastar model. Secondly
\begin{equation}
p' \approx {-\Gamma\over(r-r_p)^2} < 0.
\end{equation}
Third, assuming $\rho$ remains finite, close to the pole ($r\approx r_p$) we have
\begin{eqnarray}
 - {(\rho+p) (m+4\pi pr^3)\over r^2(1-2m/r)} \approx 
 - {[\Gamma/(r-r_p)]\;[4\pi r^3\Gamma/(r-r_p)] \over r^2(1-2m(r)/r)}
 \\
 \approx - {4\pi r_p \Gamma^2\over (1-2m(r_p)/r_p)} {1\over(r-r_p)^2} < 0.
\end{eqnarray}
So the TOV equation \emph{can} be satisfied in the vicinity of the pole provided we set
\begin{equation}
\Gamma = {1-2m(r_p)/r_p\over 4\pi r_p} > 0.
\end{equation}
That is, if the gravastar configuration is perfect fluid and finite in extent --- then this pole is the only way the TOV equation can be satisfied. (It is easy to check that higher-order poles do not even have this nice property of being compatible with the isotropic TOV equation.) Now despite the fact that the presence of a simple pole in the pressure is compatible with the isotropic TOV equation, we must reiterate that such a pressure pole is unphysical because it is a naked singularity ---  the physically correct deduction form this analysis is that gravastar-like objects must violate pressure isotropy.

\section{Bounds on the pressure anisotropy}

Once we accept that perfect fluid spheres are not what we are looking for to model gravastars, one might wonder what happens to the Buchdahl--Bondi bound for isotropic fluid spheres. It has been shown that for $\rho' < 0$ and $p_t \leq p_r$ the $8/9$ bound still holds. However if the transverse stress is allowed to exceed the radial stress, $p_t > p_r$, then the upper limit shifts to $2M/R < \kappa \leq 1$, where $\kappa$ depends on the magnitude of the maximal stress anisotropy \cite{Guven:1999wm}.\footnote{For related work on anisotropic stars, see for instance~\cite{Herrera}.}
In the gravastar picture, we shall soon see that in the crust $p_t > p_r$, and that the compactness of a gravastar is not limited by the Buchdahl--Bondi bound, but only by the magnitude of the maximal pressure anisotropy and the regularity of the metric, \emph{i.e.} $2m(r)/r < 1$. Let us now make these qualitative considerations more quantitative.
To do this, let us
rewrite \eref{eq:TOV2} in the form
\begin{equation}
\label{eq:anisotropy}
\Delta \equiv \frac{p_t - p_r}{\rho} = \frac{r}{2} \left[
\frac{p_r'}{\rho} + \left(1+\frac{p_r}{\rho} \right)\, g \right]
\, .
\end{equation}

In section \ref{sec:iso_fail} we have determined the smallest interval in radii for which
anisotropic pressure is necessary to be $(r_g,r_\mathrm{max}]$. We can now ask for explicit bounds on
$\Delta$ in that interval. By inserting the definition of $g$,
\eref{eq:TOV2}, into \eref{eq:anisotropy}, we get
\begin{equation}
\Delta = \frac{r}{2} \left[ \frac{p_r'}{\rho} +
\left(1+\frac{p_r}{\rho} \right)\, \frac{m+4\pi p_r\, r^3}{r^2
\left[ 1-2m/r \right] } \right] \, .
\end{equation}
For the interval $r\in [r_0,r_\mathrm{max}]$,  by making use of $p_r' \geq 0$ and $p_r \geq 0$, we find the simple lower bound
\begin{equation}
\label{E:ineq}
\Delta \geq \frac{1}{4}\, \frac{2m/r}{1-2m/r} > 0.
\end{equation}
Now in the region $[r_0,r_\mathrm{max}]$ we have $p_t>p_r\geq0$, consequently if the DEC is to be satisfied we must at the very least have $\Delta \leq 1$.  But this is guaranteed to be violated whenever $2m/r>4/5$. 

That is: If the gravastar is sufficiently close to forming a horizon, in the sense that $2m/r > 4/5$ somewhere in the range $[r_0,r_\mathrm{max}]$, then the DEC must also be violated at this point.\footnote{And even if we were to discard the entire positive-pressure region $(r_0,R)$, we can nevertheless still apply this bound at $r_0$ itself: If $2m(r_0)/r_0>4/5$ then the DEC is violated at $r_0$.} Consequently, any gravastar that is sufficiently close to forming a horizon will violate the DEC in its ``crust''.

For the interval
$r\in (r_g,r_0)$ we find the considerably weaker bounds
\begin{equation}
0 \leq \frac{r\,p_r'}{2\,\rho} < \Delta < \frac{r\,p_r'}{2\,\rho}
+ \frac{1}{4}\, \frac{2m/r}{1-2m/r}
\end{equation}
where we have used $p_r' \geq 0$, $p_r < 0$, and the NEC. 

\section{Minimizing the anisotropic region}

\begin{figure}[htb]
\begin{center}
\input{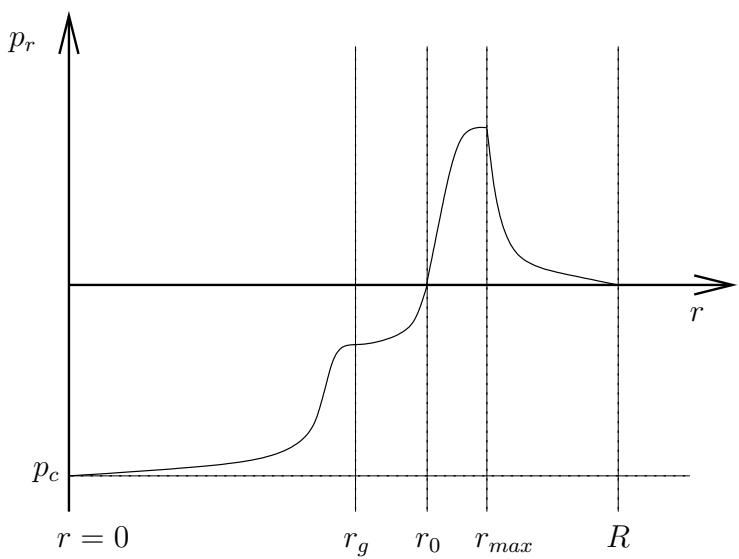}
\end{center}
\caption{\label{F:2}
Qualitative sketch of radial pressure as a function of $r$ for a gravastar that is as near as possible to a perfect fluid. Note the inflexion point at $r_g$ and the kink at $r_\mathrm{max}$.}
\end{figure}

Let us now attempt to minimize the region over which anisotropy is present. It is easy to see that at $r_g$ we have
\begin{equation}
\Delta(r_g) = \frac{r_g\,p_r'}{2\,\rho} \geq 0 \, .
\end{equation}
Therefore, in the case where the anisotropy is confined to the
smallest interval possible, we want $p_r'(r_g) = 0$, corresponding to an inflexion point for the radial pressure. At the point $r_0$ of zero
radial pressure, the anisotropy \emph{cannot} vanish:
\begin{equation}
\Delta(r_0) = \frac{r_0\,p_r'}{2\,\rho} + \frac{1}{4}\,
\frac{2m/r_0}{1-2m/r_0} > 0 \, .
\end{equation}
At the point of maximal radial pressure, the anisotropy also has
to be non-zero, at least if we take the limit from below:
\begin{equation}
\Delta(r_\mathrm{max}^-) = \frac{1}{4}\, \frac{2m/r_\mathrm{max}}{1-2m/r_\mathrm{max}}
\left(1+\frac{p_r}{\rho} \right)\, \left(1+\frac{4\pi p_r\, r_\mathrm{max}^3}{m
} \right) \, > 0 \, .
\end{equation}
Beyond the peak $\forall r > r_\mathrm{max}$,  it is possible to arrange $\Delta = 0$, though at a price:  If
we wish to confine the anisotropy to the smallest interval
possible, we have to set $\Delta(r \rightarrow r_\mathrm{max}^+) = 0$
which leads to a discontinuity in $p_r'$ and $\Delta$ as well as a
``kink'' in the pressure profile $p_r$ at $r_\mathrm{max}$. (However, $p_r$ itself is still continuous, as is the density $\rho$.) Indeed we then have
\begin{equation}
p_r'(r_\mathrm{max}^+) = 
{2 \rho_\mathrm{max}\over r_\mathrm{max}} \; \Delta(r_\mathrm{max}^-).
\end{equation}
The
implications of confining the pressure anisotropy to the smallest
interval possible are shown in figure \ref{F:2}, where the anisotropy is confined to the region $r\in(r_g,r_\mathrm{max}]$.

\def\from{\leftarrow}

The Mazur--Mottola model is now recovered as the limiting case where
$r_g\to r_0\from r_\mathrm{max}$, and so all the important anisotropy is confined in their inner
thin shell. The anisotropy
$\Delta\to\infty$, because $p'\to\infty$. Effectively $\Delta$ is replaced by 
choosing an appropriate finite (in
this case negative) surface tension $\vartheta$ and surface energy
density $\sigma$ which is given by the Israel--Lanczos--Sen 
junction conditions~\cite{visser04, Israel}. 

The second, outer thin shell which is present in the Mazur--Mottola model
is \emph{not} a physical necessity, but is merely a convenient
way to avoid an infinitely diffuse atmosphere that would
arise otherwise from the equation of state $p=\rho$. A finite
surface radius $R$ can be modelled by altering the equation of
state slightly to include a finite surface density $\rho_S$ which is
reached for vanishing pressure: $\rho(p)=\rho_S+p$. Then, the
outer thin shell can be omitted when joining the gravastar metric
onto the Schwarzschild exterior metric.

\section{Features of the anisotropic equation of state}

In the case of a perfect fluid, the geometry is completely defined
by the set of differential equations \eref{eq:mass},
\eref{eq:TOV} plus initial conditions and an equation of
state, which is commonly written as $\rho=\rho(p)$ or equivalently $p=p(\rho)$.

Once we accept the need to abandon the notion of a perfect fluid, we have to
replace the isotropic TOV equation \eref{eq:TOV} with
its anisotropic counterpart \eref{eq:TOV2} and thereby
introduce an additional (free) function $\Delta(r)$. To close the
set of equations, it is now necessary to define \emph{two} equations of
state. Naively, one might choose $p_r(\rho)$ and $p_t(\rho)$, which is a rather strong assumption that forces $p_r$ and $p_t$ to change in lock-step.
Alternatively, one could simply 
postulate a density profile $\rho(r)$ (which is the strategy adopted, for example, in
\cite{Dymnikova}) or equivalently a pressure profile, plus one equation of state.
One could also (rather unphysically) choose to specify any two profiles $\rho(r)$, $p_r(r)$, and $p_t(r)$ by hand, and use the anisotropic TOV to calculate the remaining profile.
Yet another
possibility to obtain a well defined solution lies in finding an
additional (differential) equation, that might be motivated by some 
appropriate variational principle, for example by minimizing the
``total anisotropy''
\begin{equation}
  \frac{1}{R} \int_0^R (\rho\Delta)^2 \d r 
\quad\Rightarrow\quad 
\delta\left[\int_0^R (\rho\Delta)^2 \d r \right] = 0\, .
\end{equation}
No matter whether it is an equation of state for the transverse
pressure or a variational principle for the total anisotropy, the
extra equation should (if the gravastar model is to be even qualitatively correct)  be responsible for maintaining stability of
the gravastar over a wide range of total masses and central pressures. In other words, the
closed set of equations must be self-regulating if it is to be physically interesting ---  the gravastar should shift to a new  stable
configuration when the total mass changes. 

The  central point of this article is that if you wish to believe in gravastars you must accept that there will be regions of anisotropic pressure where $\Delta > 0$. This suggests that it might be most efficient to choose the two distinct ``equations of state'' as being equations for $\rho$ and for $\Delta$. But what variables should these equations of state depend on? An obvious candidate is the radial pressure, but in view of the inequality \eref{E:ineq} it is clear that the gravastar material, if it is to succeed in generically avoiding the formation of horizons should be sensitive to the ``compactness'' $2m(r)/r$. That is, we should posit equations of state of the form
\begin{equation}
\rho= \rho(r,p_r,2m/r); \qquad \Delta= \Delta(r,p_r,2m/r);
\end{equation}
and (making everything as explicit as possible) solve the paired differential equations
\begin{eqnarray}
{\d p_r(r)\over \d r} &=& 
 -\frac{[\rho(r,p_r(r),2m(r)/r)+p_r(r)]\;[2m(r)/r+8\pi \;p_r(r) \;r^2 ]}{2\,r\;[1-2m(r)/r]} 
\nonumber \\
 &&
+{2\;\rho(r,p_r(r),2m(r)/r)\;\Delta(r,p_r(r),2m(r)/r)\over r};
\end{eqnarray}
\begin{equation}
{\d m(r)\over \d r}  = 4\pi \;\rho(r,p_r(r),2m(r)/r) \;r^2.
\end{equation}
Whether or not the gravastar model ultimately succeeds in its goals depends on whether or not one can find physically realistic equations of state that have the effect of bounding $2m(r)/r < 1$ for large ranges of $p_c$ and total mass. 

Now there is an important issue of principle here:  Traditional relativists are somewhat nervous when the variable $2m/r$ enters the equation of state, arguing that $2m/r$ is not detectable by local physics, and that an equation of state that depends on $2m/r$ somehow violates the Einstein equivalence principle. This is not correct for the following reason: $2m/r$ is certainly measurable by \emph{quasi-local effects} in small but \emph{finite size regions}. To see this note that for any static spherically symmetric spacetime the orthonormal components of the Riemann tensor are (see, for instance, \cite[p 110]{book}):
\begin{equation}
R_{\hat t \hat r \hat t \hat r} = {4\pi\over3}\; [3(\rho-p_r+2p_t)-2\bar\rho];
\end{equation}
\begin{equation}
R_{\hat t \hat \theta \hat t \hat \theta} = {4\pi\over3}\; [3p_r +\bar\rho];
\end{equation}
\begin{equation}
R_{\hat r \hat \theta \hat r \hat \theta} = {4\pi\over3}\; [3\rho-\bar\rho];
\end{equation}
\begin{equation}
R_{\hat \theta \hat \phi \hat \theta \hat \phi} = {4\pi\over3}\; [2\bar\rho].
\end{equation}
The point here is that the Riemann tensor is certainly measurable in finite-sized regions, so in particular $\bar\rho$ is measurable. Likewise $r$ is measurable in finite-sized regions, and therefore $2m/r= (8\pi/3)\;\bar\rho\;r^2$ is measurable. Consequently the compactness $2m/r$ is a quasi-local measurable quantity and it can meaningfully be put into the equation of state without violating the equivalence principle. (In either the Mazur--Mottola scenario~\cite{MM1, MM2, MM3} or the Laughlin \emph{et al.} scenario~\cite{Laughlin1, Laughlin2} the gravastar material is assumed to be a quantum condensate, and therefore sensitive to non-local physics. The point of the current discussion is that we do not need to appeal to quantum non-locality to get $2m/r$ into the equation of state --- properly understanding the equivalence principle is enough.)
Note that while this argument demonstrates that possible equations of state are at least conceivable, that is not the same as explicitly demonstrating that such equations of state actually exist. That is a challenge which we leave for the future.

\section{Results and Discussion}

In this article we have delineated the qualitative features one would expect from a gravastar configuration that avoids delta function transition layers and has finite regular pressure profile at all locations. We have used these qualitative features to place constraints on the anisotropy parameter $\Delta$, demonstrating that perfect fluid gravastars are a lost cause, and extracting some generic information regarding what the gravastar equation of state should be. 

Specifically, finite-sized  gravastar-like objects (with the key defining feature being negative central pressure) must exhibit anisotropic stresses in their ``crust'', which is the region where the pressure is increasing as one moves outwards but the local force of gravity is inwards.  Trying to build a perfect fluid gravastar  results either in an infinite-size infinite-mass object, or in a naked singularity as the pressure exhibits a simple pole. Assuming the WEC, the magnitude of anisotropy required in the crust can be explicitly bounded in terms of the local compactness $2m(r)/r$, and becomes arbitrarily large for gravastars that are sufficiently close to forming a horizon, in which case the DEC must be violated. Consequently, if one demands that the equation of state for gravastar matter results in configurations that are horizon-avoiding for large ranges of total mass and central pressure, then it is difficult to avoid the conclusion that the equation of state must depend on the local compactness $2m(r)/r$. This is certainly an unusual equation of state, but we emphasise that (when properly understood) this does not violate the equivalence principle.

While we are personally agnostic as to the existence or non-existence of gravastars, we feel it is important to understand what their general properties might be in order to have a clear understanding of what the observational evidence regarding astrophysical black holes is actually telling us.

\section*{Acknowledgements}

This research was supported by the Marsden Fund administered by the
Royal Society of New Zealand.

\section*{References}




\begin{thebibliography}{69}


\bibitem{Marek}
M.~A.~Abramowicz, W.~Kluzniak and J.~P.~Lasota,
  ``No observational proof of the black-hole event-horizon,''
  Astron.\ Astrophys.\  {\bf 396} (2002) L31
  [arXiv:astro-ph/0207270].


\bibitem{MM1}
 P.~O.~Mazur and E.~Mottola,
 ``Gravitational condensate stars,''
  arXiv:gr-qc/0109035.
  
 \bibitem{MM2}
  P.~O.~Mazur and E.~Mottola,
  ``Dark energy and condensate stars: Casimir energy in the large,''
  arXiv:gr-qc/0405111.
  
  
 \bibitem{MM3}
P.~O.~Mazur and E.~Mottola,
  ``Gravitational vacuum condensate stars,''
  Proc.\ Nat.\ Acad.\ Sci.\  {\bf 111}, 9545 (2004)
  [arXiv:gr-qc/0407075].
  
  
\bibitem{Laughlin1}
G.~Chapline, E.~Hohlfeld, R.~B.~Laughlin and D.~I.~Santiago,
  ``Quantum phase transitions and the breakdown of classical general
  relativity,''
  Int.\ J.\ Mod.\ Phys.\ A {\bf 18} (2003) 3587
  [arXiv:gr-qc/0012094].
  
 \bibitem{Laughlin2}
 G. Chapline, R. Laughlin, amd D. Santiago,
 ``Emergent relativity and the physics of black hole interiors'',
 published in \emph{Artificial black holes}, 
 edited by Mario Novello, Matt Visser, and Grigori Volovik.
 (World Scientific, Singapore, 2002).

\bibitem{Gliner}
E.~B.~Gliner,
``Inflationary universe and the vacuumlike state of physical medium,''
Phys.\ Usp.\  {\bf 45} (2002) 213
[Usp.\ Fiz.\ Nauk {\bf 45} (2002) 221].
\\
E.~B.~Gliner,
``Scalar black holes,''
arXiv:gr-qc/9808042.


\bibitem{Dymnikova}
  I.~Dymnikova and E.~Galaktionov,
  ``Stability of a vacuum nonsingular black hole,''
  arXiv:gr-qc/0409049.
\\
 I.~Dymnikova,
``Spherically symmetric space-time with the regular de Sitter center,''
Int.\ J.\ Mod.\ Phys.\ D {\bf 12} (2003) 1015
[arXiv:gr-qc/0304110].
\\
I.~Dymnikova,
``Cosmological term as a source of mass,''
Class.\ Quant.\ Grav.\  {\bf 19} (2002) 725
[arXiv:gr-qc/0112052].
\\
I.~G.~Dymnikova,
``The algebraic structure of a cosmological term in spherically symmetric
solutions,''
Phys.\ Lett.\ B {\bf 472} (2000) 33
[arXiv:gr-qc/9912116].
 \\
 I.~Dymnikova,
``Vacuum nonsingular black hole,''
Gen.\ Rel.\ Grav.\  {\bf 24} (1992) 235.
 
\bibitem{Lynden-Bell}
J. Katz and D. Lynden-Bell,
``Tension shells and tension stars'',
Classical Quantum Gravity {\bf 8} (1991) 2231-2238;
\\
Mon. Not. R. Astron. Soc. {\bf 267} (1994) 51--58.

\bibitem{Comer}
G. L. Comer and J. Katz, 
``Some conditions for the existence of tension stars'',
Mon. Not. R. Astron. Soc. {\bf 267} (1994) 51--58.
 
 
  
\bibitem{visser04}
   M.~Visser and D.~L.~Wiltshire,
  ``Stable gravastars --- an alternative to black holes?,''
  Class.\ Quant.\ Grav.\  {\bf 21}, 1135 (2004)
  [arXiv:gr-qc/0310107].
  
 \bibitem{Volovik}
 F.~R.~Klinkhamer and G.~E.~Volovik,
 ``Coexisting vacua and effective gravity,''
 arXiv:gr-qc/0503090.

\bibitem{MTW}
Charles~W. Misner, Kip~S. Thorne, and John~Archibald Wheeler.
\newblock {\em Gravitation}.
\newblock W. H. Freeman, San Francisco, 1973.

\bibitem{Rahman}
S.~Rahman and M.~Visser,
  ``Spacetime geometry of static fluid spheres,''
  Class.\ Quant.\ Grav.\  {\bf 19} (2002) 935
  [arXiv:gr-qc/0103065].

\bibitem{Martin}
   D.~Martin and M.~Visser,
  ``Bounds on the interior geometry and pressure profile of static fluid
  spheres,''
  Class.\ Quant.\ Grav.\  {\bf 20}, 3699 (2003)
  [arXiv:gr-qc/0306038].
  
  \bibitem{Bilic}
  N.~Bilic, G.~B.~Tupper and R.~D.~Viollier,
  ``Born-Infeld phantom gravastars,''
  arXiv:astro-ph/0503427.
 
 
  
 \bibitem{Harrison}
 B.K. Harrison, K.S. Thorne, M. Wakano, and J.A. Wheeler, 
 \emph{Gravitation theory and gravitational collapse}, 
 (University of Chicago Press, Chicago, 1965).
 
 
 \bibitem{Yunes}
  M.~Visser and N.~Yunes,
``Power laws, scale invariance, and generalized Frobenius series: Applications
to Newtonian and TOV stars near criticality,''
Int.\ J.\ Mod.\ Phys.\ A {\bf 18} (2003) 3433
[arXiv:gr-qc/0211001].
  

\bibitem{Guven:1999wm}
  J.~Guven and N.~O'Murchadha,
  ``Bounds on 2m/R for static spherical objects,''
  Phys.\ Rev.\ D {\bf 60}, 084020 (1999)
  [arXiv:gr-qc/9903067].
  
  \bibitem{Herrera}
  L. Herrera and N.O.Santos,
  ``Local anisotropy in self-gravitating systems'', 
  Phys. Rep. {\bf 286} (1997) 53. 
  \\
   L.~Herrera, A.~Di Prisco, J.~Martin, J.~Ospino, N.~O.~Santos and O.~Troconis,
  ``Spherically symmetric dissipative anisotropic fluids: A general study,''
  Phys.\ Rev.\ D {\bf 69}, 084026 (2004)
  [arXiv:gr-qc/0403006].
  \\
  L.~Herrera, J.~Martin and J.~Ospino,
  ``Anisotropic geodesic fluid spheres in general relativity,''
  J.\ Math.\ Phys.\  {\bf 43}, 4889 (2002)
  [arXiv:gr-qc/0207040].
  \\
  L.~Herrera, A.~Di Prisco, J.~Ospino and E.~Fuenmayor,
  ``Conformally flat anisotropic spheres in general relativity,''
  J.\ Math.\ Phys.\  {\bf 42}, 2129 (2001)
  [arXiv:gr-qc/0102058].
    
  
  \bibitem{Israel}
  W.~Israel,
  ``Singular Hypersurfaces And Thin Shells In General Relativity,''
  Nuovo Cim.\ B {\bf 44S10} (1966) 1
  [Erratum-ibid.\ B {\bf 48} (1967\ NUCIA,B44,1.1966) 463].
  \\
  C.~Barrabes and W.~Israel,
  ``Thin shells in general relativity and cosmology: The Lightlike limit,''
  Phys.\ Rev.\ D {\bf 43} (1991) 1129.
  \\
  K. Lanczos,
``Fl\"achenhafte Verteilung der Materie in der Einsteinschen Gravitationstheorie'', 
Ann. Phys. (Leipzig) {\bf 74} (1924) 518--540.
\\
N. Sen,
``\"Uber die Grenzbedingungen des Schwerefelds an unstetigen 
Kreisfl\"achen'',
Ann. Phys. (Leipzig) {\bf 73} (1924) 365--396.

 
  \bibitem{book}
  Matt Visser, \emph{Lorentzian wormholes: from Einstein to Hawking},
  (AIP Press, New York, 1995).
  
  
  
\end{thebibliography}
\end{document}